\documentclass[sigconf]{acmart}
\usepackage{geometry}
\usepackage[utf8]{inputenc}
\usepackage{xspace}
\usepackage{url}
\usepackage{bm}
\usepackage{bbm}
\usepackage{subcaption}

\newcommand{\yes}{\textsc{yes}\xspace}
\newcommand{\no}{\textsc{no}\xspace}
\usepackage{amsfonts, amsmath, amsthm}
\usepackage{macros}
\usepackage{mathtools}
\usepackage{hyperref}
\usepackage[utf8]{inputenc}
\usepackage{color}

\theoremstyle{definition}

\usepackage{xcolor}
\usepackage[draft,inline,nomargin,index]{fixme}

\fxsetup{theme=color,mode=multiuser,inlineface=\itshape,envface=\itshape}
\FXRegisterAuthor{sv}{asv}{\colorbox{gray!10!white}{\color{black}Suresh}}
\FXRegisterAuthor{ah}{aah}{\colorbox{blue!10!white}{\color{black}Aaron}}
\FXRegisterAuthor{mg}{amg}{\colorbox{blue!10!white}{\color{black}Marissa}}
\FXRegisterAuthor{mm}{amm}{\colorbox{green!10!white}{\color{black}Malika}}
\FXRegisterAuthor{kk}{akk}{\colorbox{gray!10!white}{\color{black}Kweku}}
\fxusetargetlayout{color}

\title{The Misuse of AUC: What High Impact Risk Assessment Gets Wrong}
\author{Kweku Kwegyir-Aggrey}
\affiliation{
\institution{Brown University}
\city{Providence}
\country{USA}
}
\email{kweku@brown.edu}
\author{Marissa Gerchick} 
\affiliation{
\institution{American Civil Liberties Union}
\city{New York, NY}
\country{USA}
}
\email{mgerchick@aclu.org}

\email{}
\author{Malika Mohan}
\affiliation{
\institution{American Civil Liberties Union}
\city{New York, NY}
\country{USA}
}
\email{mmohan@aclu.org}

\author{Aaron Horowitz}
\affiliation{
\institution{American Civil Liberties Union}
\city{New York, NY}
\country{USA}
}
\email{ahorowitz@aclu.org}

\author{Suresh Venkatasubramanian}
\affiliation{
\institution{Brown University}
\city{Providence}
\country{USA}
}
\email{suresh@brown.edu}
\date{November 2020}

\begin{abstract}
    When determining which machine learning model best performs some high impact risk assessment task, practitioners commonly use the Area under the Curve (AUC) to defend and validate their model choices.  In this paper, we argue that the current use and understanding of AUC as a model performance metric misunderstands the way the metric was intended to be used.  To this end, we characterize the misuse of AUC and illustrate how this misuse negatively manifests in the real world across several risk assessment domains.  We locate this disconnect in the way the original interpretation of AUC has shifted over time to the point where issues pertaining to decision thresholds, class balance, statistical uncertainty, and protected groups remain unaddressed by AUC-based model comparisons, and where model choices that should be the purview of policymakers are hidden behind the veil of mathematical rigor. 
    We conclude that current model validation practices involving AUC are not robust, and often invalid. 
\end{abstract}
\copyrightyear{2023} 
\acmYear{2023} 
\setcopyright{rightsretained} 
\acmConference[FAccT '23]{2023 ACM Conference on Fairness, Accountability, and Transparency}{June 12--15, 2023}{Chicago, IL, USA}
\acmBooktitle{2023 ACM Conference on Fairness, Accountability, and Transparency (FAccT '23), June 12--15, 2023, Chicago, IL, USA}
\acmDOI{10.1145/3593013.3594100}
\acmISBN{979-8-4007-0192-4/23/06}

\begin{document}

\maketitle

\section{Introduction}
\label{sec:intro}
Predictive models are being developed as \emph{risk assessment tools} in use cases that range from credit \cite{NCLC_Report}, to housing \citep{rodriguez2020all}, medicine \cite{Verbakel2020-kg}, criminal justice \cite{PATTERNus2019first, Terranova_undated-pq, northpointe2022compas}, and more. In these cases, the predictive model is described by a score function that outputs a \emph{risk score} for an individual, and a threshold such that scores above the threshold are deemed to be ``high risk.'' In many cases, there are multiple thresholds to divide risk scores into different categories of risk, corresponding to different levels of intervention.

The predictive models underlying these risk assessments are often validated using a specific measure of performance known as the Area Under the Curve (AUC). The AUC attempts to provide an aggregate measure of performance of the model across many different thresholds, and is used in two ways. Firstly, the AUC is used as an absolute measure of model quality -- a model with an AUC above some cutoff is considered admissible. Secondly, the AUC is used to compare different models -- one model is considered superior to another if its AUC is higher. Typically, model selection happens in two steps: first, the AUC is used to select a score function, and then thresholds are selected based upon the chosen score function. 

The AUC plays a critical role, not just in the design of risk assessments, but in their defense.
In settings such as the criminal legal system, child welfare, or education, the scientific validity of the choice of particular risk assessments are debated (in court, and in the public sphere) by reference to the AUC \cite{loomis-opinion, dieterich2016compas, vaithianathan2019implementing, bruch2020using}.\footnote{For example, in \textit{Wisconsin v. Loomis}, arguments about the COMPAS risk assessment's ``accuracy'' relied on AUC \cite{loomis-opinion}, and COMPAS's developers focused significantly on AUC in their response to ProPublica's reporting \cite{angwin2016machine} about racial bias and COMPAS's outputs \cite{dieterich2016compas}.} Debates about the choices of thresholds used are often in reference to a model that was selected using the AUC. Thus, the AUC represents a baseline validation choice that all further contestation must take for granted.

\paragraph{Summary of Contributions}

Our goal in this work is to argue that the AUC as currently used is not a valid measure of goodness for risk assessment tools. We start with a three-part critique. First, we reexamine the rationale for the use of AUC in model selection (as a replacement for model accuracy evaluation) and show that it is an inherently probabilistic and noisy measure whose interpretation depends heavily on properties of the data being used for evaluation, as opposed to its current use as a single data-independent number. We derive this interpretation from a careful analysis of works in statistics and machine learning that have studied the AUC over the past few decades, but whose messages have been lost in the move from the academic literature to deployment. Second, we show that the use of AUC for model (and threshold) selection -- through the process of first choosing a model and then choosing thresholds -- masks a deeper domain-specific and policy discussion around how to weigh different kinds of errors a model may make and their consequences for individuals subjected to the model. 
Third, we demonstrate the use of AUC for multi-level risk assessments is mathematically incoherent, and its use in the context of ``fair'' risk assessments that seek to mitigate racial bias is also meaningless.
Finally, we couple our critique with with an evaluation of how model selection is performed in three domains of relevance: the criminal legal system, the child welfare system, and in education. In all of these domains, we demonstrate that the use of AUC to justify model selection and deployment is deeply flawed, and identify several examples of risk assessments used by government agencies that rely on AUC.  

We conclude by exhorting model developers and practitioners to consider sociotechnical context, select appropriate metrics, and recognize the policy implications of their choices. We suggest actionable takeaways for these audiences, including a more context-sensitive choice of evaluation metric, a careful and holistic examination of model performance, and continuous monitoring and re-evaluation. 


\section{Technical Background}
\label{sec:background}

In this section, we define AUC in the context of risk assessments and related background. A reader familiar with the AUC may skip this section on a first reading. 

\paragraph{Decision Procedures And Error Measurement}

The starting point for defining AUC is a decision procedure that takes an input and returns a binary output that is either \yes or \no. The question being asked can vary across different forms of risk assessments. However, we will in general assume a question of the form "Is an individual at risk?" (of failing to appear in court, being arrested for a crime, being investigated for alleged child neglect, or dropping out of school, and so on). In other words, we will assume that the \yes answer is the one that corresponds to the event the decision-maker ostensibly seeks to avoid. Formally, we will denote the decision procedure (the risk assessment tool) by a decision function $D_\lambda(x)$ that is made up of a \emph{score function} $r(x)$ that outputs a non-negative number and a \emph{threshold parameter} $\lambda$ that takes features $x$ describing an individual and produces one of $\{\yes, \no\}$. The score function $r(x)$ is usually trained on historical data of the form $\{x_i, y_i\}$ where $x_i$ is an individual and $y_i$ represents the decision made for them by some past process. The decision $D_\lambda(x)$ is defined as \looseness-1
\[ D_\lambda(x) = \begin{cases} 
            \yes & r(x) \ge \lambda \\ 
            \no & \text{otherwise} 
        \end{cases}\]
 A single number that describes the performance of such a decision function is its \emph{predictive accuracy}:
\begin{align*}
\frac{1}{n}\sum_{i=1}^{n} \mathbbm{1}\{{D_{\lambda}(x_i) = y_i}\}
\end{align*}

A more nuanced description of performance is a \emph{confusion matrix} (see Table~\ref{tab:confusion}) that tabulates the number of correct classifications (true positives (TP) and true negatives (TN)) and incorrect classifications (false positives (FP) and false negatives (FN)). The subscript $\lambda$ serves to reinforce that these fractions depend on the threshold parameter. 
\begin{table}[htbp]
\centering
\begin{tabular}{lrr} \toprule
   &$ y_i = \yes$& $y_i = \no$\\ \midrule
  $D_\lambda(x_i) = \yes$ & $\text{TP}_{\lambda}$ & $\text{FP}_{\lambda}$\\
  $D_\lambda(x_i) = \no$ & $\text{FN}_{\lambda}$ & $\text{TN}_{\lambda}$ \\ \bottomrule
\end{tabular}
\caption{Confusion matrix for a binary decision procedure}
\Description{A table displaying the confusion matrix for a binary decision procedure. Specifically, when both the decision function and label are YES, we call this a true positive.  When both the decision function and label are NO then we call this a true negative.  Similarly, when the decision function outputs YES for an instance labeled NO we call this a false positive, and when the decision function outputs NO for an instance labeled YES we call this a false negative.}
\label{tab:confusion}
\end{table}

\paragraph{AUC}
\label{sec:auc}
Consider what happens when we fix the score function $r(x)$ but let the threshold $\lambda$ increase from $0$. 
Let the \emph{true positive rate} be the fraction of $\yes$ inputs for which $D_\lambda(x) = \yes$. This can be written as 
$\text{TPR}_\lambda = \frac{\text{TP}_\lambda}{\text{TP}_\lambda + \text{FN}_\lambda}$.
Similarly, let the \emph{false positive rate} be the fraction of $\no$ inputs for which $D_\lambda(x) = \yes$. This can be written as 
$\text{FPR}_\lambda = \frac{\text{FP}_\lambda}{\text{FP}_\lambda + \text{TN}_\lambda}$.
As $\lambda$ varies, we can plot $(\text{FPR}_\lambda, \text{TPR}_\lambda)$ as a point in two dimensions. If we do this for all values of $\lambda$ we get a plot that is often called the Receiver Operating Characteristics (ROC) curve.\footnote{Note that as described this does not need to be a curve. In practice it is merely a cloud of points since we only ever work with a finite number of parameter values.} 
In the case when we set $\lambda = 0$, $D_0(x) = \yes$ for all inputs $x$, which implies that $\text{TN}_0 = \text{FN}_0 = 0$ and corresponds to the point $(1,1)$. Similarly, if we set $\lambda = \infty$, we get no false positives and no true positives, corresponding to the point $(0,0)$. Ideally, we would  want  to achieve all true positives with as few false positives as possible, which corresponds to some point $(\text{FP}_{\lambda^*},1)$ for some threshold $\lambda^*$.

As $\text{FP}_{\lambda^*}$ gets smaller and smaller, this curve takes the shape of a vertical line $x=0$ followed by the horizontal line $y = 1$. The \emph{area under this curve} (AUC) is therefore $1$. Conversely, as the best decision process gets worse, the curve moves further and further way from the ``ideal'' point $(0,1)$ while still being anchored at $(0,0)$ and $(1,1)$, thus reducing its area. The worst possible classifier is one that merely assigns \yes and \no labels at random. For such a classifier, as we increase the parameter $\lambda$, each new point labeled \yes has an equal chance of being a true positive or a false positive, and so the curve resembles the straight line from $(0,0)$ to $(1,1)$, which has an AUC of $0.5$. 

\begin{figure}[htbp]
  \centering
  \includegraphics[width=0.45\textwidth]{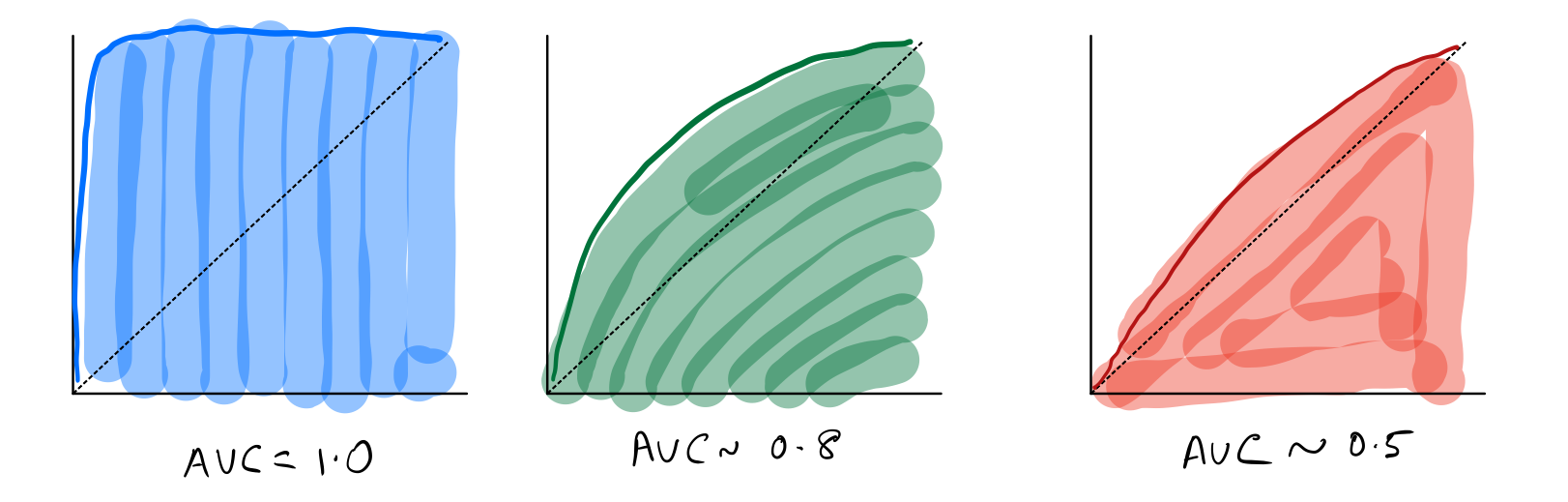}
  \caption{An illustration of AUC with different curves}
  \label{fig:auc1}
  \Description{An illustration of three different AUC curves. The area under the visualized curves decrease in AUC value from left to right.  The rightmost curve has perfect area underneath (AUC) of 1.0, and is equal to the line y=1.  The second curve resembles a sqrt(x) curve and has an AUC of 0.8.  The last curve has an AUC of 0.5 and is the line y=x from 0 to 1.}
\end{figure}

To summarize, the AUC attempts to measure the degree to which the classifier can produce a distinct decision point (represented by a single $\lambda$) that achieves a good trade-off between true and false positives. The larger the AUC, the more well defined this decision point will be.

\section{A Brief History of the AUC in Model Selection and related work}
\label{sec: related_works}
In the 1980's, Receiver Operating Characteristic (ROC) curves had long been established as a useful method for visualizing a classifier's performance and subsequently selecting a suitable decision threshold \citep{mcneil1984statisticalROC, metz1978basicRocAnalysis}.   The area under these curves was identified as a method of ``reducing[ing] an entire ROC curve to a single quantitative index of diagnostic accuracy,'' allowing for easier comparisons of different classification schemes at the model level, before thresholds come into play \citep{hanley1982meaning}. At this time, scientists viewed the AUC as a statistical property of the ROC curves, and often used statistically principled methods like hypothesis testing, in order to conduct model comparisons(using AUC as a test statistic). \citep{metz1980StatisticalSignificance, 1983RocComparisonStatistical}.   

Moving into the 90s, the use of AUC was further cemented as several works identified the measure as a convenient solution to certain issues in cost-sensitive classification at the time.  Specifically, accuracy was often criticized to its efficacy (or lack thereof) in instances of class imbalance, or its implicit assumption that all misclassifications were of equal importance \citep{swetsMeasuring1988}.  Many researchers viewed AUC as a viable alternative, due to its ability to measure a model's performance on average, without specifying misclassification costs, which were often believed to be unknown or difficult to quantify in many problem domains of the time \cite{bradley1997AUC, provost1998case, provost1997analysis}. 
Shortly thereafter, there was a sudden proliferation in efforts that championed the AUC over accuracy as the preferred metric and comparator of model performance \citep{ling2003auc, cortes2003auc, 2000robustAUC}.  It was at this time that the meticulous statistical procedures that previously characterized AUC-based comparisons were disregarded, and AUC was no longer purely viewed as a statistical property of the ROC curve, but rather as ``provably superior to the empirical misclassification rate as an evaluation measure'' \cite{rosset2004AUC}.\looseness-1  

With that said, the ``superiority'' of AUC, or any performance metric for that matter, remains a debated topic.  Arguments against the AUC as a measure of model performance have been previously made, for example, in \citet{hand2009measuring, adamsComparing1999}.  The majority of these arguments call into question, according to \citet{lobo2008auc}, the ``reliability [of AUC] as a comparative measure of accuracy between model results.''   
 Additionally, several studies have analyzed the centrality of AUC in risk assessments more broadly \citep{bokhari2014dangerous, bokhari2023clinical, Helmus2017-ui, fazel2022predictive, Singh2013-kk}. \citet{fazel2022predictive} conducted a systematic review of validation studies of risk assessments used at sentencing, finding that the sole performance metric reported in most studies was AUC. \citet{Singh2013-kk} examined studies of risk assessments that attempt to predict risk of violent recidivism from the 1980s, 1990s, and early 2000s, finding that AUC is often misinterpreted in these studies and that many studies present conflicting validation benchmarks.

 Additional consideration to the metric was given by \citet{kallus2019fairnessXauc} and \citet{narasimhan2020pairwise} who propose a fair rendition of AUC.   Previous works considered other AUC variants such as an extension to support multi-class classification \citep{hand2001simple}, or Partial AUC \citep{dodd2003partial}.  We also do not focus on AUC maximization in this work, given that we are primarily evaluating the AUC in model evaluation and not in model training.  For a survey of AUC maximization literature we reference the survey work of \citet{yang2022auc}.

\section{Comparing Models using AUC is Harder Than it Looks}
\label{sec:comparing_models_using_auc}
The goal of this section is to highlight three key issues in AUC based model comparison which are often ignored in risk assessment. First, in Section
\ref{sec: AUC_accuracy_disagree} we dispute the common assumption that AUC can be used to measure predictive accuracy.  In Section \ref{sec:probability} we explain how the two metrics are correlated, and discuss the context dependent factors which can alter the strength and reliability of their correlation --  without taking these factors into account however, AUC based measurements of predictive accuracy can be noisy and inaccurate.  In cases where AUC is \textit{not} a noisy signal for accuracy, the situation hardly improves.  In Section \ref{sec: auc_obfuscates:threshold_decisions} we show that AUC based model comparisons obscure the specific impact of decision thresholds which are needed to operationalize the risk assessment model.  As a matter of fact, we emphasize that each decision threshold can be linked to a normative assessment on the cost of an incorrect classification, e.g., how detrimental is an incorrect pretrial decision.  Interrogating the impact of these thresholds requires analyzing a classifier's ROC curve, which is impossible to recover from AUC alone. Unfortunately, ROC analysis is not a panacea either.  In Section \ref{sec: multiclass_fairness_issues} we describe that in most risk assessment settings ROC curve analysis and best-case AUC comparisons are still insufficient to ensure ``good'' risk assessment.  Indeed, risk assessments models are often used to wager predictions across multiple risk groups, or multiple protected groups.  In either case, both the AUC and the ROC curve are poor signals of model performance on a specific group -- particularly when such a group is small.  All in all, we argue that AUC alone offers model comparisons that are serendipitiously correct at best, but in most cases typical of risk assessment, provably brittle.\looseness-1

\subsection{Characterizing the Relationship Between AUC and Predictive Accuracy}
\label{sec:auc_accuracy_relationship}
Current model comparison practices rely on the AUC as a measure of model performance under the assumption that higher AUC implies higher model accuracy.  In this section we will demonstrate that this assumption is wrong, and the relationship between the two metrics is significantly more complex than this assumption suggests. 

\subsubsection{Classifiers Rank Differently Under AUC and Accuracy} 
\label{sec: AUC_accuracy_disagree}
The assumption that higher AUC implies higher accuracy is easy to disprove with examples -- in fact there are many situations where a model can have can have higher AUC than another model but worse predictive accuracy, or vice versa.  To make this clear, we use the following rank-based, but equivalent definition of AUC.   Let $r_i$ be the rank of all positive examples (examples labeled $\yes$) under some classifier, then the AUC can be written as 
\begin{align*}
    AUC = \frac{S - n_\no(n_\no + 1)/2}{n_\yes n_\no} \quad \text{with} \quad S = \sum r_1 + r_2 + \dots r_{n_\yes}
\end{align*}

Consider two classifiers $A,B$ whose classifications of the same 10 $\no$ and $\yes$ elements are written in rank order as shown in Figure \ref{table: high_AUC_worse_accuracy} -- although classifier $B$ has the worse AUC ($0.64$ versus $0.84$), it has better accuracy ($80\%$) than classifier $A$ ($60\%$) at the shown (and common) threshold of $0.5$. This observation is not affected by a shift in threshold. If we were to move the threshold of classifier $A$ one position to the left, its accuracy would increase to 70\% without changing its AUC,\footnote{Recall that AUC does not depend on the decision threshold.} but B still has better accuracy than A.  This observations remains true even if we choose the accuracy-maximizing threshold for each classifier separately. In Figure \ref{table: high_AUC_worse_maxaccuracy}  the classifiers $C,D$ are classifying the same elements as $A$ and $B$ but with decision thresholds chosen for each classifier to maximize classifier accuracy. As before, classifier $D$ has the worse AUC but the higher accuracy.

\begin{figure}[htp]

\begin{subfigure}{\columnwidth}
\begin{tabular}{lllllllllll}
Classifier A & - & - & - & + & \multicolumn{1}{l|}{+} & - & - & + & + & + \\ 
Classifier B & + & - & - & - & \multicolumn{1}{l|}{-} & + & + & + & + & - \\
\end{tabular}
\caption{Classifier $A$ has AUC 0.84 and Accuracy 60\%; Classifier $B$ has AUC $0.64$ and Accuracy 80\%} 
\label{table: high_AUC_worse_accuracy}

\end{subfigure}

\bigskip

\begin{subfigure}{\columnwidth}
\begin{tabular}{lllllllllll}
Classifier C & - & - & - & + & \multicolumn{1}{l|}{-} & + & - & + & + & + \\ 
Classifier D & - & - & - & \multicolumn{1}{l|}{-} & + &  + & + & + & - & + \\
\end{tabular}
\caption{Classifier $C$ has AUC 0.88 and Accuracy 80\%; Classifier $D$ has AUC $0.84$ and Accuracy 90\%}
\label{fig:example_table}
\end{subfigure}

\caption{Two scenarios demonstrating that higher AUC does not imply higher accuracy when comparing models}
\label{table: high_AUC_worse_maxaccuracy}
\Description{The figure consists of two tables, where each table contains two rows.  In each row is a ranked list of positive and negatively labeled examples. Underneath each table is their AUC and accuracy computations. The goal of the figure is to dispute the assumption that high AUC implies higher accuracy.}
\end{figure}

\begin{figure}
\centering

\end{figure}

\begin{figure}
\centering

\end{figure}  

These examples illustrate the fundamental distinction between AUC and accuracy.  The AUC is a measure of how well \yes and \no are ranked, whereas accuracy measures correctness of $\yes$ or $\no$ predictions with respect to some threshold.  Depending on the specific classification task, these two notions could yield vastly different conclusions about model behavior.  

What then is the real relationship between AUC and accuracy? We explore this relationship next: it turns out that while there is a connection between the two, the two measurements cannot be linked deterministically; instead we ought to reason about their relationship using probability. 

\subsubsection{The Relationship Between AUC and Accuracy is Statistical and Data Dependent}
\label{sec:probability}
The probabilistic relationship between between AUC and accuracy is best understood through the following equivalent definition of the AUC.   
If we let $X$ be the score distribution of the $\yes$ labeled elements and $Y$ be the distribution of the $\no$ labeled elements, it is well known that
\begin{align}
    AUC = \Pr[X \geq Y]. 
\end{align}
Without perfect knowledge of the $X$ and $Y$ distributions, it is impossible to compute this probability -- or equivalently the AUC -- exactly.  In settings when these distributions are unknown, the AUC can be (and is often) estimated from samples.  This implies that prior stated formulations of the AUC are actually estimators of the \emph{empirical} AUC, whereas the \emph{true} AUC is a property of the score distributions for the positive and negative classes.  

Following this reasoning, any empirically observed AUC is subject to the same statistical rules that govern any other random variable, e.g., we can estimate its mean, it is subject to some variance, etc.  For this reason it will be helpful to consider what AUC values are possible for a given classifier, or in other words, how can we determine the distribution of AUC values for a given classifier?  This distribution, as described in \citep{cortes2003auc} depends most crucially on three parameters: an error rate, the ratio of $\yes$ and $\no$ instances, and the sample size.   

\begin{description}
\item[Error Rate $\epsilon$.] We define the error rate as the fraction of elements which are misclassified.  For small error rates, the distribution of AUC values tends towards higher values that are well concentrated around their mean.  For large error rates however, AUC values tend to be lower on average, but also highly variable.
\item[Sample Size  $n = n_\yes + n_\no$.] When $n$ is large we generally can estimate the distribution of AUC values with high confidence.  When $n$ is small, observed AUC values may exhibit more variance than they would in a high sample regime.
\item[Class Balance $k = \frac{n_\no}{n_\yes+n_\no}$.]  When classes are highly imbalanced, AUC values exhibit large variance.  When classes are balanced, AUC values exhibit small variance, meaning AUC and accuracy correlate more predictably.   
\end{description}

From these quantities, we can describe exactly how empirical AUC values may vary on a given sample. To reason about this distribution of possible values, we'll describe the average AUC of this distribution, denoted $AUC_{avg}$, and its standard error, denoted $SE(AUC)$.  Like \citet{cortes2003auc}, we assume all misclassifications are equally probable.  Thus, the average AUC as a function of the error rate $\epsilon$ can be interpreted as the most likely empirical AUC value for a classifier that makes $n_{err}$ miclassifications, where $\epsilon = \frac{n_{err}}{n_\no + n_\yes}$. Consider the following formula for $\AUC_{avg}$ from \citet{cortes2003auc}
\begin{align}
\label{eq: auc_mean}
1 - \epsilon - \frac{(n_\no-n_\yes)^2(n_\no + n_\yes + 1)}{4 n_\no n_\yes}\left (\epsilon - \frac{\sum_{\ell=1}^{n_{err}-1}{n_\no + n_\yes \choose \ell}}{\sum_{\ell=0}^{n_{err}}{n_\no + n_\yes + 1 \choose \ell}}\right) 
\end{align}
In the special case when classes are perfectly balanced $k=\frac{1}{2}$ (or $n_\no = n_\yes$) , the expected AUC value for a classifier only depends on its error rate, i.e. $\AUC_{avg}=1-\epsilon$. \footnote{This average equally weights all classifications/rankings with the given error rate.} Generally speaking however, classes are seldom balanced; this is typical of high impact risk assessment where there are elevated stakes associated with the assessment task and the underlying event is a rare occurrence (see Section \ref{sec:gov-examples-rai} for examples of this imbalance). In the cases where $k \gg \frac{1}{2}$, AUC values may be low on average, despite having a classifier with relatively low error.\footnote{We remind the reader that in unbalanced settings, low error can still be indicative of poor performance on the \yes class. See \citep{bradley1997AUC}.} We visualize the interplay between class balance and error rate in Figure \ref{figure:auc_error_table}.  

\begin{figure*}
\begin{tabular}{clllllllllllllll}
\cline{1-1} \cline{3-16}
\begin{tabular}[c]{@{}c@{}}Class Balance \\ $k$ \end{tabular} & \multicolumn{1}{c}{} & \multicolumn{1}{c}{} & \multicolumn{1}{c}{} & \multicolumn{1}{c}{} & \multicolumn{1}{c}{} & \multicolumn{1}{c}{} & \multicolumn{1}{c}{} & \multicolumn{2}{c}{error ($\epsilon$)} & \multicolumn{1}{c}{} & \multicolumn{1}{c}{} & \multicolumn{1}{c}{} & \multicolumn{1}{c}{} & \multicolumn{1}{c}{} & \multicolumn{1}{c}{} \\ \cline{1-1} \cline{3-16} 
 &  & \multicolumn{1}{c}{0\%} & 2.5\% & \multicolumn{1}{c}{5\%} & 7.5\% & 10\% & 12.5\% & 15\% & 17.5\% & 20\% & 22.5\% & 25\% & 27.5\% & 30\% & 32.5\% \\
0.50 &  & 1.000 & 0.980 & 0.950 & 0.920 & 0.900 & 0.880 & 0.850 & 0.820 & 0.800 & 0.780 & 0.750 & 0.720 & 0.700 & 0.680 \\
0.55 &  & 1.000 & 0.980 & 0.950 & 0.920 & 0.900 & 0.870 & 0.850 & 0.820 & 0.790 & 0.770 & 0.740 & 0.720 & 0.690 & 0.660 \\
0.60 &  & 1.000 & 0.970 & 0.950 & 0.920 & 0.890 & 0.860 & 0.830 & 0.810 & 0.780 & 0.750 & 0.720 & 0.690 & 0.660 & 0.630 \\
0.65 &  & 1.000 & 0.970 & 0.940 & 0.910 & 0.880 & 0.850 & 0.810 & 0.780 & 0.750 & 0.710 & 0.680 & 0.640 & 0.600 & 0.560 \\
0.70 &  & 1.000 & 0.970 & 0.930 & 0.890 & 0.860 & 0.820 & 0.780 & 0.740 & 0.700 & 0.660 & 0.610 & 0.560 & 0.500 & {\color[HTML]{656565} 0.450} \\
0.75 &  & 1.000 & 0.960 & 0.920 & 0.870 & 0.830 & 0.780 & 0.730 & 0.680 & 0.620 & 0.570 & 0.500 & {\color[HTML]{656565} 0.430} & {\color[HTML]{656565} 0.360} & {\color[HTML]{656565} 0.270} \\
0.80 &  & 1.000 & 0.950 & 0.890 & 0.830 & 0.770 & 0.710 & 0.650 & 0.570 & 0.500 & {\color[HTML]{656565} 0.430} & {\color[HTML]{656565} 0.330} & {\color[HTML]{656565} 0.230} & {\color[HTML]{656565} 0.120} &  \\
0.85 &  & 1.000 & 0.930 & 0.850 & 0.770 & 0.690 & 0.600 & 0.500 & {\color[HTML]{656565} 0.400} & {\color[HTML]{656565} 0.290} & {\color[HTML]{656565} 0.180} & {\color[HTML]{656565} 0.040} &  &  &  \\
0.90 &  & 1.000 & 0.890 & 0.760 & 0.630 & 0.500 & {\color[HTML]{656565} 0.370} & {\color[HTML]{656565} 0.210} & {\color[HTML]{656565} 0.040} &  &  &  &  &  &  \\ \hline
\end{tabular}
\caption{Recall $n_\no$ is the number of elements in the negative class, and $n_\yes$ the positive class. For a given level of class imbalance $k = \frac{n_\no}{n_\no+n_\yes}$ this table shows the expected AUC, as a function of the error rate $\epsilon$.  Our computation of expected AUC assumes all misclassifications are equally likely.  Observed values for empirical AUC may differ, depending on total sample size $n = n_\no+no_\yes$. Entries in gray correspond to AUC < 0.5, which is performance worse than a classifier which predicts each class with 50\% odds.}
\label{figure:auc_error_table}
\end{figure*}
\begin{table}[]
\Description{A table showing the relationship between average case AUC, class balance, and error. The goal of the table is to demonstrate that the average AUC typically increases with the error rate of the classifier, however the strength of this correlation depends on the balance of positive and negative examples.  A highly imbalanced sample suggests a very weak link between AUC and accuracy. Conversely, a highly balanced sample implies a strong correlation.}
\end{table}

\begin{figure}
\centering
\includegraphics[width=\columnwidth]{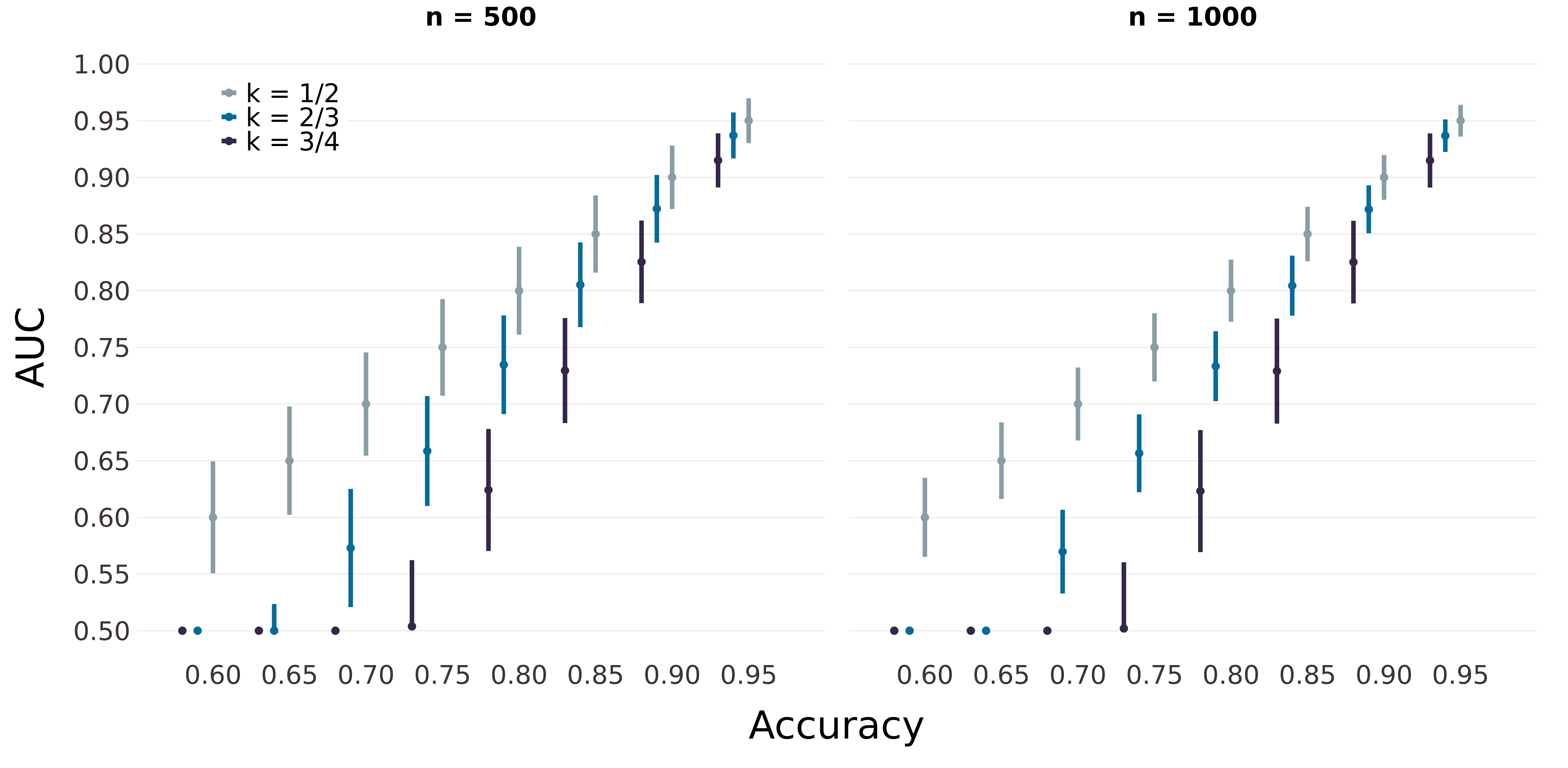}
\caption{95\% Confidence intervals for the AUC based on classifier accuracy for n=500 (left) and n = 1000 (right).  Mean and standard error computed using Eq. \ref{eq: auc_mean} and Eq. \ref{eq:auc_se} respectively.}
\label{fig:auc_intervals}   
\Description{A color plot visualizing ninety-five percent confidence intervals for AUC.  On the x-axis is classifier accuracy, on the y-axis is AUC.  The confidence intervals on the figure are shown for three different levels of class imbalance.  The figure corroborates the table above it, and shows that class imbalance heavily deteriorates the positive correlation between AUC and accuracy, and that this effect becomes more severe as a classifier becomes more inaccurate.}
\end{figure}

Similarly, the tightness/variability of any AUC distribution also crucially depends on $k, \epsilon$ but as a function of $\AUC$.  We use the following formula for standard error (of AUC) from \citet{hanley1982meaning}.  Let $\theta$ be some estimate of the true AUC (we used $\theta = \AUC_{avg}$) then $SE(AUC)$ is exactly, 
\begin{align}
\label{eq:auc_se}
     \sqrt{\frac{\theta(1-\theta) + (n_\yes - 1)(Q_1 - \theta^2) + (n_\yes - 1)(Q_2 - \theta^2)}{n_\yes n_\no}} \\  \text{where} \quad Q_1 = \frac{\theta}{2 - \theta}, Q_2 = \frac{2\theta^2}{1 + \theta}
\end{align}
Classifiers with low error rates produce more consistent estimates of a classifier's true AUC.  On the other hand, classifiers with high errors rates have incredibly wide AUC distributions.   These distributions can be made even more variable when $n$ is small, or $k$ is large; large $n$ and $k=1/2$ are the least variable setting.   We show 95\% confidence intervals for the average AUC of a classifier with some setting of $k,\epsilon$ for $n=500$ and $n=1000$ in Figure \ref{fig:auc_intervals}. This figure shows how a single AUC value could be associated with a number of possible underlying classifier error rates. For example, applying a horizontal line test on the y-axis at $AUC=0.65$ indicates that this AUC value, depending on $k$ could be the result of a classifier with accuracy as low as 65\% or as high as 80\%. From this example we can see that comparing two models based on their AUC is something that must be done with care, especially in high error or imbalanced class regimes.\looseness-1



\begin{figure}
\centering
\includegraphics[width=.75\columnwidth]{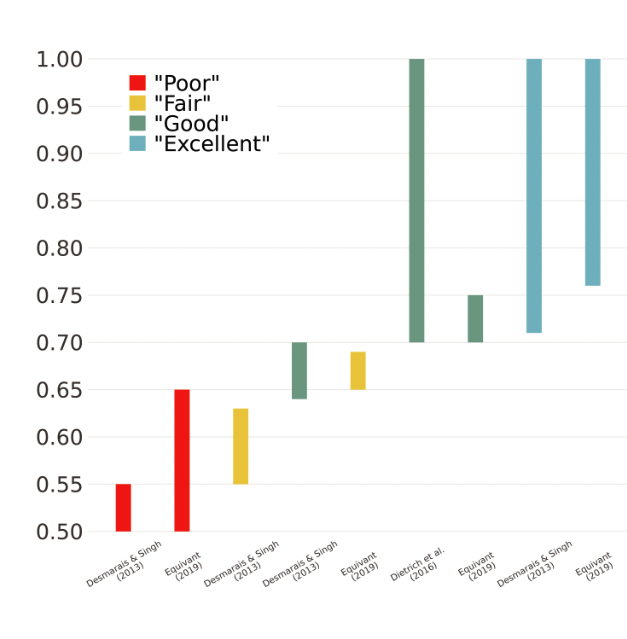}
\Description{Color coded plot comparing normative labels such as ``poor'' or ``good'' assigned to ranges of AUC values in methodology reports, showing conflicts in the labels used in different methodology reports.}
\caption{Methodology reports evaluating government uses of risk assessments present conflicting interpretations of what values constitute a ``good'' AUC.}
\label{fig:normative_auc}
\end{figure}


\subsection{AUC Obfuscates Threshold Decisions}
\label{sec: auc_obfuscates:threshold_decisions}
To describe the obfuscation of threshold decisions by AUC, we recall the following idea from cost sensitive classification literature:  in binary classification, each choice in threshold can be linked to a quantification of the cost of different types of various errors, namely false positives and false negatives.  Suppose $C_{FN}, C_{FP}$ are the costs of false negative and false positive classification.  We can compute the total cost of a classifier's errors at a threshold by computing
\begin{align}
\label{eq: cost_loss}
    \Lagr_{\cost}(\lambda) = n_{\no}C_{FN}(1-TPR_\lambda) + n_{\yes}C_{FP}FPR_\lambda. 
\end{align}
In the special case $C_{FN} = C_{FP}$, this expression measures the number of misclassifications at $\lambda$.  

To find a threshold which minimizes $\Lagr_{\cost}(\lambda)$ we can refer to the ROC curve. Recall, that we can parameterize ROC curve over thresholds, i.e., $ROC(\lambda) = (TPR_\lambda, FPR_\lambda)$.  The optimal threshold for a given choice  $C_{FN}, C_{FP}$ is the threshold on the ROC curve, which has slope proportional to the the ratio of costs \citep{meekins2018cost}. Formally, if $\lambda$ is optimal for $C_{FN}, C_{FP}$ then 
\begin{align}
\label{eq:roc-deriv}
    \frac{d}{d\lambda}ROC(\lambda) = \frac{n_{\yes} C_{FP}}{n_{NO}C_{FN}}
\end{align}
From this equality, we learn two key ideas: (1) Any choice in threshold implies an assessment of the relative merits of false positives and false negatives, i.e., specific values of misclassification costs.  Although declaring these costs explicitly is difficult in a practical sense \citep{stevenson2022pretrial}, we can interpret the costs implied by a threshold as a framework for quantifying the presumed cost of incorrectly assessing risk;  (2) The ROC curve can be used to compute optimal thresholds (with respect to Eq. \ref{eq: cost_loss}), and for this reason is a useful utility in navigating  model performance for various misclassification costs, or equivalently, various decision thresholds \cite{fawcett2004roc}.\looseness-1 

It is this recognition of the utility of ROC curves, which reveals another pitfall of the AUC.  By definition as an integral over the ROC curve, the AUC mathematically discards threshold level information, specifically the derivatives in the l.h.s. of Eq \ref{eq:roc-deriv}.  By only comparing AUCs without comparing the ROC curve from which they came, practitioners lose the ability to evaluate the consequences of misclassifications across  thresholds.  Unfortunately, it is this exact evaluation of costs which many have argued is necessary to maintain the integrity and accountability of automated decision making pipelines \cite{corbett2017algorithmic, stevenson2022pretrial}.   Additionally, we remind the reader that it is possible for a model to have higher AUC but worse performance at the thresholds which are of interest (see Fig \ref{fig:example_table}). In consideration of these facts, we conclude that AUC's inability to compare model performance at specific thresholds is a serious shortcoming of the metric. 
\subsection{Issues in Fairness and Multi-Class Classification}
\label{sec: multiclass_fairness_issues}
A common goal of risk assessment is to stratify individuals into different semantically meaningful risk categories, e.g, low, medium, and high.  To accomplish this, practitioners often take a learned model $r$ and simply apply \textit{multiple} decision thresholds. Suppose these thresholds $\lambda_1 < \lambda_2 < \dots < \lambda_{k-1}$ are provided, where
\begin{align}
\label{eq: multiclass}
     D(x) =
  \begin{cases}
    O_1 & r(x) < \lambda_1 \\
    O_2 & \lambda_{1} \le r(x) < \lambda_2 \\
    \dots & \dots \\
    O_k & r(x) \ge \lambda_{k-1} 
  \end{cases}
\end{align}

We say that $r(x)$ is well fitted if the output of $r$ accurately predicts an individual's true underlying risk of attaining the \yes outcome.  A crucial issue with the AUC is that it does not measure model fit \citep{lobo2008auc}, and as a result can often lead to inaccurate risk predictions for when there are more than two risk groups.  We can illustrate this point with an example.\looseness-1
\begin{example}
Let $r$ be some scoring function that is well fitted, and let $r'(x) = \frac{r(x)}{2}$ be a new classifier which is the output of $r$ times $\frac{1}{2}$. The following two illustrations in Figure~\ref{fig: risk_vs_fit} show that while predictive accuracy is damaged by tampering with the output of $r$ through this transformation, the AUC remains the same, and thus cannot help us distinguish between a model with high predictive accuracy and one with low accuracy. 
 
\begin{figure*}[h!]
 \caption{A comparison of two classifiers $r$ and $r'$ with $\AUC(r) = \AUC(r') = 1$ but different predictive accuracy across 3 risk groups. The location of a shape represents $r(x_i)$. Circles denote an instance labeled $\no$ and squares denote an instance labeled $\yes$.  The color of each shape represents the ground truth level of risk: green is for a low risk instance, yellow for medium risk, and red for high risk. As in Eq. \ref{eq: multiclass}, there are 3 outcomes groups: the instances to the left of $\lambda_1$ receive decision $O_1$, the instances between $\lambda_1$ and $\lambda_2$ receive $O_2$, and the instances to the right of $\lambda_2$ receive $O_3$. }
    \centering
    \begin{subfigure}[t]{0.49\textwidth}
        \centering
        \includegraphics[scale=.23]{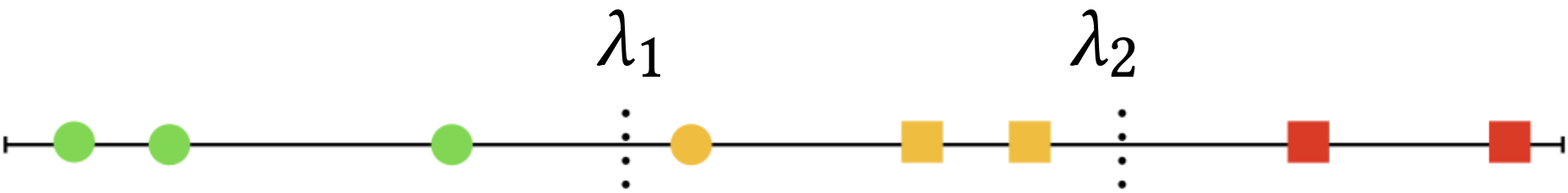}
        \caption{The output of $r(x_i)$ on several elements.}
    \label{subfig: g_output}
    \end{subfigure}%
    ~
    \begin{subfigure}[t]{0.49\textwidth}
        \centering
        \includegraphics[scale=.23]{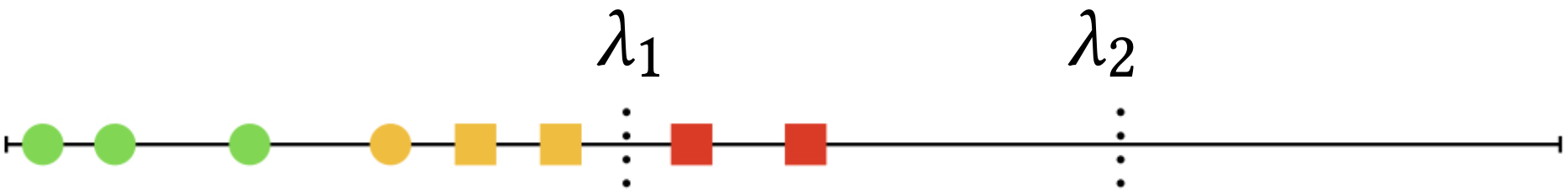}
        \caption{The output of $r'(x_i) = r(x_1)/2$ on several elements.}
    \label{subfig: gp_output}
    \end{subfigure}
\label{fig: risk_vs_fit}
\Description{A comparison of two classifiers $r$ and $r'$ with $\AUC(r) = \AUC(r') = 1$ but different predictive accuracy across 3 risk groups. The location of a shape represents $r(x_i)$. Circles denote an instance labeled $\no$ and squares denote an instance labeled $\yes$.  The color of each shape represents the ground truth level of risk: green is for a low risk instance, yellow for medium risk, and red for high risk. As in Eq. \ref{eq: multiclass}, there are 3 outcomes groups: the instances to the left of $\lambda_1$ receive decision $O_1$, the instances between $\lambda_1$ and $\lambda_2$ receive $O_2$, and the instances to the right of $\lambda_2$ receive $O_3$.}
\end{figure*}
In Fig. \ref{subfig: g_output} (left) each outcome group corresponds to the correct level of risk.  In Fig. \ref{subfig: gp_output}, the first risk group contains a mixture of low and medium risk elements, while the second risk group now captures high risk.  In both cases, all positive elements are ranked above negative elements, indicating that both $r$ and $r'$ are perfect at \emph{discriminating} between positive and negative elements (i.e. have $\text{AUC}=1$) however $r'$ is not \emph{well fitted}, given that low medium and high risk scores are fairly close together. \looseness-1
\end{example}
 
This example illustrates that the AUC cannot measure how well a model is fit, which is critical to multi-class classification problems.  Indeed, if the model is not well fit, the application of various decision thresholds can lead to disparate accuracy for the different groups -- be those risk groups, or protected groups.\footnote{We explore this through several real world examples in Section \ref{sec:gov-examples-rai}.}   Even worse, as we highlight in Section 5, practitioners often compare the AUC across groups, claiming that if a model has equal or similar AUC performance on various groups, the tool is neutral with regard to group-related biases.  This specific claim is disputed by \citet{kallus2019fairnessXauc} who show that the inability of AUC to measure model fit means that equal intra-group AUC is not sufficient to assert a model's fairness or groupwise neutrality. 


\subsection{Takeaways: Contextualizing AUC's Problems in Risk Assessment}
\label{sec:takeaways}

We conclude this section by placing the above issues in the context of risk assessment, briefly summarizing the problems that arise when using AUC in connection with evaluating risk assessments. 

\subsubsection*{Takeaway \#1 (Section \ref{sec:auc_accuracy_relationship}) - Although one model may have higher AUC than another, there is a nonzero probability that the first model may perform worse (in terms of accuracy) than the model with lower AUC.  Furthermore, correctly executing a model comparison with AUC requires considering context dependent factors, which are frequently ignored in current risk assessment validation.}  As we see in Fig \ref{fig:auc_intervals}, classifiers with high error rates can have a wide range of possible AUC values, so using AUC as a proxy for accuracy can lead to incorrect conclusions some of the time.  In summary, AUC values above .75 tend to reflect low classifier error and exhibit little variance. AUC values in the 0.6 to 0.75 range, which are typical of risk assessment tools used in criminal justice \cite{Singh2013-kk}, are associated with high classifier error rate and are significantly more variable, making AUC a noisy signal of class accuracy; this noise is more prominent when classes are imbalanced (also a common characteristic of risk assessments).\looseness-1 


\subsubsection*{Takeaway \#2 (Section \ref{sec: auc_obfuscates:threshold_decisions}) - AUC alone cannot evaluate the performance of a model once a decision is applied even though thresholded decisions reflect key policy choices and expectations of a risk assessment model, and are crucial to its successful deployment.}   By definition, the AUC measures model performance before the application of a threshold.  As a consequence, it is often the case where a model has higher AUC, but worse performance for certain decision thresholds.  The choice in threshold is linked to the goals of the risk assessment tool, but this choice is mathematically overlooked by AUC. \looseness-1

\subsubsection*{Takeaway \#3 (Section \ref{sec: multiclass_fairness_issues}) - Models validated using AUC often have different accuracy across risk groups and protected groups} Often, risk assessment tools look to allocate interventions across a number of risk categories, e.g., low, medium, high.  Although a binary classification model can easily be extended to a multi-class problem, the AUC cannot be extended comparably without modification (see \citet{hand2001simple}). Similar discrepancies occur across protected groups -- AUC validation on its own is insufficient to claim that a model performs equally well across protected groups \citep{kallus2019fairnessXauc}.   

\section{Use of AUC in Risk Assessments}
\label{sec:gov-examples-rai}

Misunderstandings of AUC are not merely statistical or theoretical concerns. When government agencies procure or build tools based on AUC-based claims, these practices can impact whether people involved with the criminal legal system are incarcerated pretrial, whether families will be investigated by child welfare agencies, whether students will be labeled as ``at-risk'' in school, and more. 

Using examples from the criminal legal system, the child welfare system, and education, we now analyze how government agencies rely on AUC to justify the use of risk assessment tools, advocate for their continued use, and argue that they are not racially biased. As we will show, government agencies often treat AUC as an objective measure of predictive accuracy -- even in the common scenario of extreme class imbalance in validation samples -- and the use of AUC in these contexts is often predicated on the problematic and inaccurate assumptions about AUC discussed in the previous sections. As highlighted in \cite{Singh2013-kk} and further in Figure \ref{fig:normative_auc}, the inconsistencies of AUC-related practices also extend to how government agencies characterize success when it comes to AUC. Government uses of risk assessment instruments often present absolute -- yet conflicting -- normative evaluations of what constitutes a good AUC, sometimes referencing the normative ranges provided in \citet{desmarais2013risk}. While some discussions of this topic note that normative labels of AUC ranges should be context-specific \cite{fazel2022predictive, Helmus2017-ui, compascore2019}, these labels sometimes conflict even within specific domains such as the criminal legal system (see Figure \ref{fig:normative_auc}).\footnote{While not the main focus of this work, we note that these kinds of normative labels are not just limited to government uses of risk assessment. For example, a 2021 report by the company HireVue stated that ``AUC values above .60 suggest the model is able to distinguish between two classes fairly well'' \cite[p.~13]{hirevueAUC}.}  

\subsection{Risk Assessment in the Criminal Legal System}
\label{sec:crim-leg}

Risk assessment instruments are currently used in many states throughout the U.S. at various stages of the criminal legal system. These assessments are used to inform pretrial release decisions, sentencing decisions, decisions about the services that will be made available to people who are incarcerated, and more \cite{PATTERNus2019first, Terranova_undated-pq, northpointe2022compas}. Researchers, civil rights organizations, community members, policymakers and others have extensively criticized and debated these tools and the manner in which they contribute to racial bias in the criminal legal system \cite{eckhouse2019layers, angwin2016machine, corbett2017algorithmic, appeal-rat}. For many of these risk assessments, AUC plays an important role in the tool's creation and use. In this section, we'll primarily discuss a criminal risk assessment tool called the ``Prisoner Assessment Tool Targeting Estimated Risk and Needs'' or \emph{PATTERN}. We claim that PATTERN's use is predicated on many of the problems summarized in Section \ref{sec:takeaways} and contextualize these issues through an analysis of PATTERN's threshold cutoffs, noting the clear policy implications of the threshold setting.

PATTERN is a risk assessment tool currently used by the U.S. Department of Justice to evaluate people incarcerated in federal prisons. The tool, which was developed in connection with the First Step Act of 2018 \cite{PATTERNus2019first}, attempts to estimate the likelihood that a person will be rearrested or returned to federal custody following release from federal prison. Since the first version of PATTERN was developed in 2019, the tool has been substantively altered several times (as of writing, PATTERN 1.3, the third version of the tool, is currently in use). PATTERN 1.3 has separate male and female assessments and also produces two separate risk scores, one for ``general recidivism'' and one for ``violent recidivism.'' Thresholds are applied to each score to group incarcerated people into ``risk level categories,'' and for each person, these ``risk level categories'' based on the two different risk scores are aggregated. PATTERN's outputs directly determine some of the resources individuals are able to access while incarcerated and their ability to earn credits for programming that can be applied towards earlier release \cite{pattern2021report}. Civil rights organizations, researchers, impacted community members and other stakeholders have repeatedly criticized PATTERN since it was developed, highlighting numerous serious issues with the tool’s design, development, and use. These issues include PATTERN's reliance on data that reflects the racial discrimination of the criminal legal system, problematic use of the tool at the onset of the COVID-19 pandemic, and serious errors with both the technical systems and human processes used to calculate PATTERN scores -- errors that have led to miscalculated risk scores for tens of thousands of people \cite{pattern2019letter, paipatterncritique, brennancenterPATTERN, pattern2022comment}. 

AUC has been a key metric underpinning PATTERN since the tool was developed in 2019. While not the only metric used to evaluate PATTERN, AUC has been used to justify the tool’s creation, to inform model development, and to make arguments about PATTERN's ``racial neutrality.'' While the use of AUC as a validation metric is not inherently bad, in the context of PATTERN, AUC has been used to make conclusions about the tool in ways that suffer from the issues discussed in Section \ref{sec:comparing_models_using_auc}. For example, a 2019 development report describing the first version of the tool states that the research team ``rel[ied] on the AUC as the primary metric for evaluating predictive validity'' \cite[p.~50]{PATTERNus2019first}. The report compares AUC values for PATTERN to reported AUC for other criminal risk assessment instruments, concluding that ``PATTERN
achieves a higher level of predictability and
surpasses common risk assessment tools for
correctional population in the U.S.'' \cite[p.~56]{PATTERNus2019first}. In one telling visualization, the tool's creators declare that PATTERN is ``15\% more predictive'' than other criminal risk assessment tools because its AUCs are roughly 15\% higher than the average of the reported AUCs of several other criminal risk assessment tools \cite[p.~57]{PATTERNus2019first} -- a conclusion that is unsupported on the basis of a mere AUC comparison alone, as discussed in Section \ref{sec:comparing_models_using_auc}.

As we discuss in Section \ref{sec:auc_accuracy_relationship}, it is important to note the level of class imbalance associated with these results. For the PATTERN general recidivism tool, the sample was roughly balanced -- 45.4\% of individuals in the sample experienced a return to custody or a rearrest within three years of release from incarceration. Conversely, for the violent recidivism tool, the sample was highly imbalanced (15\% of individuals were rearrested for a suspected act of violence within three years of release from incarceration) \cite[p.~15]{pattern2021report}. Importantly, under current DOJ practice, an individual's score on the violent recidivism tool overrides their score on the general recidivism tool \cite[p.~13]{fsa2022report}, so the class imbalance in the violent recidivism sample (and the resulting effects on interpreting the AUC) has the potential to affect PATTERN as a whole  upon deployment. For the violent recidivism tool, with a reported AUC of .77 - .78 on this sample \cite[p.~57]{PATTERNus2019first}, the underlying classifier, in the average case, has error around 7.5\% (see Table \ref{figure:auc_error_table}). Note that while this error rate may seem benign, it does not tell the full story: this error rate could be achieved by misclassifying up to 50\% of the people rearrested for a suspected act of violence.\footnote{A 50\% error rate on 15\% of the sample creates an overall error rate of 7.5\%.} 

In its most recent validation reports, the Department does acknowledge some of the limitations of AUC, stating that AUC is ``less useful for gauging the accuracy of a risk tool as used in practice. For instance, the AUC does not provide information about the accuracy of the high-risk [risk-level category] designation'' \cite[p.~22]{pattern2021report}. Though the Department makes this acknowledgement and includes some threshold-specific evaluation metrics in recent reports about PATTERN, it still relies on AUC to justify the tool's continued use and to 
examine PATTERN's ``racial and ethnic neutrality,'' grouping by race and comparing AUC values in recent validation reports \cite{pattern2021report, pattern2022report}, a practice also employed in prior validations of PATTERN \cite{pattern2020revalidation, pattern2020update}. For example, a 2020 report stated that, with regard to racial bias, ``PATTERN is a neutral assessment tool, as evidenced by the nearly equal scores [for different racial groups] on the Area Under the Curve (AUC) analysis'' \cite[p.~9]{pattern2020update} -- a statement that, as highlighted in Section \ref{sec: multiclass_fairness_issues}, reflects an inaccurate interpretation of AUC.

As highlighted in Section \ref{sec: auc_obfuscates:threshold_decisions}, AUC does not evaluate the performance of a model at specific thresholds -- but thresholds are crucial to PATTERN's deployment, and setting thresholds for a tool like PATTERN is arguably an important question of policy. The First Step Act of 2018 requires that PATTERN classify each incarcerated person as either minimum, low, medium, or high risk for recidivism \cite{PATTERNus2019first}. The Department of Justice has repeatedly altered the thresholds that define these risk categories. The thresholds for the earliest version of PATTERN were developed by relying on the base rate of recidivism (re-arrest or return to custody), using different thresholds for the male and female scales and the ``general recidivism'' and ``violent recidivism'' scales \cite[p.~50]{PATTERNus2019first}. In subsequent validation reports, the Department describes procedures for repeatedly changing these cut points for various reasons \cite{pattern2020update, pattern2020revalidation, pattern2021report} including to ``help mitigate the effects of various racial and ethnic disparities associated with previous risk groupings'' for PATTERN 1.3 \cite[p.~14]{fsa2022report}. This acknowledgement -- that adjusting the thresholds is partly an attempt to address racial bias -- is a clear indication that threshold-setting is a policy choice.\footnote{As highlighted in other analyses of risk assessment instruments used in the criminal legal system \cite{eckhouse2019layers}, intuitive interpretations of labels like ``high risk'' may also be misleading. For example, only 35\% of individuals in PATTERN's training data who would have been classified as ``high risk'' on the male violent recidivism instrument under PATTERN 1.3 were re-arrested or returned to BOP custody \cite[p.~20]{pattern2021report}.} 

Flawed interpretations of AUC for model comparison and racial bias analyses underpin PATTERN. More broadly, the use of AUC does not evaluate a crucial aspect of PATTERN's deployment -- threshold-setting -- which implicates important questions of policy. PATTERN is not the only criminal justice risk assessment tool that relies on AUC in this manner; several additional examples are summarized in Table \ref{tab:other-gov-examples}.

\subsection{Risk Assessment in the Child Welfare System}
\label{sec:child-welfare}

Although child welfare agencies in the U.S. are increasingly using predictive tools in their decision-making processes \cite{samant2022family}, many organizations have raised concerns about these tools and their use in child welfare systems marked by entrenched discrimination \cite{abdurahman2021calculating, glaberson2019coding, saxena2020human, eubanks2018automating, williamsfamilyregulation,  stapleton2022imagining}.\footnote{While we use the term ``child welfare system'' here for clarity, we note that some scholars characterize this system as one of family policing or family regulation rather than one focused on child welfare \cite{roberts2009shattered, roberts2022torn, williamsfamilyregulation, hinaHRWreport}.} These types of tools are used by child welfare agencies at various decision-making points, including informing screening decisions, where call screening workers at child welfare agencies evaluate allegations of neglect and decide whether to investigate those allegations. In this context, risk assessments are sometimes used to estimate the likelihood that a child will be removed from their home within a certain time period following a referral to a child welfare agency \cite{samant2022family}.  Here, we'll describe the use of AUC in connection with this type of risk assessment deployed in Los Angeles County, California \cite{la_childwelfare_tool}, Douglas County, Colorado \cite{vaithianathan2019implementing}, and Allegheny County, Pennsylvania \cite{vaithianathan2019allegheny}, connecting these uses to some of the issues described in Section \ref{sec:comparing_models_using_auc}. 

The use of each of these risk assessments relies on AUC as an important metric (though not the only metric) in model development and evaluation, often using samples characterized by severe class imbalance. For the LA Risk Stratification Tool \cite{la_childwelfare_tool}, developed in the last several years and recently piloted in several areas of Los Angeles County, AUC was used for model selection \cite[p.~13]{la_childwelfare_tool} and to evaluate the model's fairness by calculating race-specific AUC values \cite[p.~15]{la_childwelfare_tool}. The reported AUC value for the chosen model was .83, and the sample was highly imbalanced (13.8\% of the sample experienced removal from their homes) \cite[p.~15]{la_childwelfare_tool}). For the Allegheny Family Screening Tool (AFST), which has been in use in Allegheny County, Pennsylvania since 2016 \cite{vaithianathan2017developing}, AUC has been used for several versions of the model to measure model fit \cite[p.~14]{vaithianathan2017developing}, perform model selection \cite[p.~7]{vaithianathan2019allegheny}, and compute group-specific AUC by race \cite[p.~14]{vaithianathan2017developing}, \cite[p.~7]{vaithianathan2019allegheny}. In one instance, the tool developers state their finding of a higher AUC (.77) for non-Black children compared to Black children (.74), ``suggest[s] that the tool was slightly better at predicting outcomes for non-black children than for black children'' \cite[p.~7]{vaithianathan2019allegheny}. For the version of the tool described in \cite{vaithianathan2019allegheny}, the overall AUC value was .76 \cite[p.~7]{vaithianathan2019allegheny} on a testing sample where approximately 1 in 8 children where removed from their homes within two years \cite[p.~7]{vaithianathan2017developing}. Lastly, for a similar tool currently in use in Douglas County, Colorado -- the Douglas County Decision Aide (DCDA) \cite{vaithianathan2019implementing} -- AUC is used to evaluate ``model performance,'' alongside other metrics. The AUC results for the considered model architectures varied from .80 to .86 \cite[p.~18]{vaithianathan2019implementing}, and in this instance, the sample used to compute these AUCs may have also been highly imbalanced. \footnote{While \cite{vaithianathan2019implementing} does not explicitly list the percentage of samples in the validation sample where a child was removed from their home, it does give a geographic breakdown of removal rates in the full training and validation sample, indicating overall removal rates in the sample were near 10\% \cite[p.13]{vaithianathan2019implementing}.} 

In these examples, the comparison of race-specific AUC to analyze algorithmic fairness may be misleading, as discussed in Section \ref{sec: multiclass_fairness_issues}. In addition, given the level of class imbalance in these instances, AUC may be a noisy signal of accuracy -- caution is warranted when interpreting AUC values as a measure of accuracy, as discussed in Section \ref{sec:auc_accuracy_relationship}. In addition, the thresholds associated with converting risk scores into risk categories for some of these tools have changed over time \cite[p.~33]{rittenhouse2022} -- important changes with policy implications that warrant close analysis.

\subsection{Risk Assessment in Education}
\label{sec:education}
Around the country, schools and school districts use ``early warning systems'' (EWS) to analyze student data and attempt to make predictions about how students will perform academically or whether they will drop out of school in the future. These systems place students into ``risk categories'' accordingly, which are shown to school administrators \cite{bowers2021early, bowers2019education, coleman2019better, balfanz2019early, doe_brief}. In 2016, these systems -- which typically use historical student data such as academic performance (e.g., grades, test scores), behavioral and attendance records, and sometimes external data --  were estimated to be in use in more than half of all public high schools in the United States \cite{doe_brief}, and recent reporting indicates at least eight state education agencies in the U.S. have built or are building algorithmic EWS \cite{feathers2023markup}. As with the examples from the child welfare system and the criminal legal system, AUC is commonly used in developing and validating EWS. But this reliance on AUC and emerging evidence about the effects of EWS raise complex questions about real-world performance, the policy choices embedded in threshold-setting, and impacts for students' lives.

A 2020 report authored by the research organization Mathematica and prepared for the Institute of Education Sciences (IES) -- an arm of the Department of Education -- provides insight into how AUC is used in the creation and deployment of EWS \cite{bruch2020using}. Using data from Allegheny County, Pennsylvania, the report analyzed machine learning approaches for building EWS that predict the likelihood of several different outcomes including: low GPA, course failure, chronic absenteeism, and suspension from school. In selecting and comparing potential models, the report relied on AUC as a measure of accuracy, writing that the ``predictive model risk scores identify at-risk students with a moderate-to-high level of accuracy,'' based on AUC values, which ranged from .75-.92 across the different outcomes and grade levels of students analyzed \cite[p.~2]{cattell2021identifying}. The report characterizes an AUC of .70 or higher as indicative of ``a strong model fit,'' referencing \citet{Rice2005-fm}, using AUC to validate model performance at the population level, and also as a benchmark to validate performance across different racial groups. One key issue affecting the interpretation of AUC values in this context is class imbalance.  Specifically, the sample used to compute AUC for several of the outcomes is imbalanced (for example, in the sample, suspension occurred less than 10\% of the time and chronic absenteeism occurred less than 25\% of the time) \cite[p.~8]{bruch2020appendix}, raising concerns about interpretations of AUC as discussed in Sections \ref{sec:auc_accuracy_relationship} and \ref{sec: multiclass_fairness_issues}. See \cite{app9153093} for further discussion of issues of class imbalance in EWS.

Beyond AUC, the report also considers other performance metrics in its analysis \cite{bruch2020using}. \citet{bruch2020using} and a related report also written by Mathematica \cite{cattell2021identifying} discuss various considerations for finding and setting thresholds to convert model predictions into risk categories for EWS, noting that the determination of cutoffs should be a context-dependent inquiry for individual school districts. \citet{cattell2021identifying} provides an example guide for use in this decision process \cite[p.~12]{cattell2021identifying}, which highlights the myriad policy implications of determining thresholds, including considerations of costs of false positives and false negatives. 

In a 2015 research paper, the developer of an EWS currently in use in schools in Wisconsin -- the Wisconsin Dropout Early Warning System (DEWS) -- made similar recommendations, including about the importance of weighing the costs of different kinds of errors based on the context in question \cite{knowles2015wisconsinEWS}. AUC was used for model selection for DEWS \cite{knowles2015wisconsinEWS}, and recent research uses AUC results to characterize the predictive value of the tool \cite{perdomo2023difficult}. But that same research also demonstrates the limits of AUC and other metrics as meaningful indicators of real world outcomes -- \citet{perdomo2023difficult} say they found that even though DEWS was accurate at predicting individual outcomes (using AUC and other metrics), it had no impact on graduation rates. \citet{perdomo2023difficult}'s findings suggest that targeting individual students with interventions without addressing the structural issues and inequities that lead students to drop out is an ineffective approach. Investigative reporting about the tool also found that DEWS had a much higher false positive rate for Black and Latino students compared to white students, that the tool's (often false) ``high-risk'' labels stigmatized students, and that educators received little training about the tool \cite{feathers2023markup}.

The impact of DEWS demonstrates that the metrics and measures used to build and measure models, including AUC, may present an incomplete picture of algorithmic performance in the real world. In deployment, understanding the nature of the services or interventions informed by model predictions, as well as the context in which the tool operates, is essential. For example, in educational contexts, Black students are disproportionately punished and subjected to violence in schools compared to white students \cite{cops-no-counselors, doe-schools-policing, beyondzero, pps-equity-stats, allegheny-school-stats}, students of color are disproportionately pushed into the criminal legal system \cite{cops-no-counselors, doe-schools-policing, homer2020police}, and surveillance technologies are increasingly being used to track and monitor students, often without their awareness or ability to opt out \cite{student-surveillance}. As educators around the country use these kinds of algorithms, context is a critical part of model evaluation and development (if algorithmic models are to be used at all). Table \ref{tab:other-gov-examples} includes more examples of EWS, and see \cite{bowers2012we, bowers2019receiver, bowers2021early, goldhaber2020assessing} for further analysis of these tools.\looseness-1 
\section{Conclusion}
\label{sec:conclusion-auc}

This paper shows that current model validation practices involving AUC are not robust, often invalid, and may hide policy choices under the guise of technical development. 
Indeed, the dominance of AUC as an evaluation metric reflects a long-standing pattern of treating the algorithms and their developers as an above-reproach black box. 
We argue the use of AUC in this manner can be considered an example of a ``framing trap'' \cite{Selbst2019-cs} persistent across domains:  model choices that should be the purview of policymakers and those directly impacted by an algorithm's use are hidden behind the veil of mathematical rigor. The framing trap also limits accountability and responsibility on the model developer's side of risk assessment design. 


This work is not a comprehensive analysis of AUC -- it is an analysis of AUC as the status quo in risk assessments used by governments, and our analysis suggests several takeaways for practitioners. First, for myriad reasons, researchers or developers of risk assessment tools must understand the context in which the tool may ultimately operate. Several recent works \cite{Abebe2020-ii, green2021data, bao1s} highlight the importance of understanding the application context as part of technical design processes. Practitioners shouldn't ignore context -- and defer decisions about thresholds, costs of errors, and associated trade-offs -- by using summary metrics like AUC. 

Informed by this context, practitioners should recognize that setting thresholds and choosing metrics for model evaluation are often choices with policy implications, and should be treated as such. Practitioners should choose the right metric(s) for the task at hand (which may or may not include AUC), and should work with relevant stakeholders in this process, considering the effects of errors for impacted communities. For example, \cite{delgado2022uncommon} discuss how computer scientists, lawyers, judges, and other researchers used an iterative co-design process to define and select evaluation metrics for legal AI applications, and in the context of pretrial risk assessments, \citet{stevenson2022pretrial} reason about threshold selection by articulating the harms of model decisions and errors. Whichever metric(s) are chosen, evaluations of the performance of the tool should realistically represent how it is deployed, considering thresholds and costs of errors. When AUC is used, care should be paid to issues of class imbalance, multi-class classification, and other topics discussed in this paper. As the example of DEWS demonstrates \cite{perdomo2023difficult}, model performance results don't tell the full story, and ``good'' performance on specific metrics isn't a panacea for issues with algorithmic tools or the systems in which they operate. Finally, given all of these dynamics, auditing of algorithmic tools should be continuous and should include regular re-evaluations of the metrics used to measure those tools. 

Although the use of AUC is a seemingly small choice among many in creating and using risk assessments, the issues that we highlight here related to the use of AUC are connected to broader concerns with risk assessments, and AUC's dominance as an evaluation metric reflects a long-standing pattern of treating the algorithms and their developers as above reproach. Governments and model developers have an obligation to set expectations about how models are developed in conversation with communities, as they would for any other policy making endeavor.

\newpage
\bibliographystyle{ACM-Reference-Format}
\bibliography{aucbib}
\newpage
\onecolumn
\section{Appendix}
\subsection{Additional Versions of Table \ref{figure:auc_error_table}}
We include additional versions of the Table \ref{figure:auc_error_table} for $n=50$ and $n=5000$ for reference; the entries for $n \gg 5000$ are the same as the $n=5000$ table -- up to 3 significant digits.  Unlike Table \ref{figure:auc_error_table}, entries less than 0.5 are omitted for clarity.  We remind the reader that these entries correspond to a classifier that is worse than random.  
\begin{table*}[h!]
\begin{tabular}{clllllllllllllll}
\cline{1-1} \cline{3-16}
\begin{tabular}[c]{@{}c@{}}Class Balance \\ $k$\end{tabular} & \multicolumn{1}{c}{} & \multicolumn{1}{c}{} & \multicolumn{1}{c}{} & \multicolumn{1}{c}{} & \multicolumn{1}{c}{} & \multicolumn{1}{c}{} & \multicolumn{1}{c}{} & \multicolumn{2}{c}{error ($\epsilon$)} & \multicolumn{1}{c}{} & \multicolumn{1}{c}{} & \multicolumn{1}{c}{} & \multicolumn{1}{c}{} & \multicolumn{1}{c}{} & \multicolumn{1}{c}{} \\ \cline{1-1} \cline{3-16} 
 &  & \multicolumn{1}{c}{0\%} & 2.5\% & \multicolumn{1}{c}{5\%} & 7.5\% & 10\% & 12.5\% & 15\% & 17.5\% & 20\% & 22.5\% & 25\% & 27.5\% & 30\% & 32.5\% \\
0.50 &  & 1.000 & 0.980 & 0.960 & 0.920 & 0.900 & 0.880 & 0.840 & 0.820 & 0.800 & 0.780 & 0.760 & 0.720 & 0.700 & 0.680 \\
0.55 &  & 1.000 & 0.980 & 0.959 & 0.919 & 0.899 & 0.878 & 0.838 & 0.817 & 0.797 & 0.776 & 0.756 & 0.715 & 0.694 & 0.674 \\
0.60 &  & 1.000 & 0.978 & 0.957 & 0.913 & 0.891 & 0.869 & 0.825 & 0.802 & 0.780 & 0.757 & 0.734 & 0.687 & 0.663 & 0.639 \\
0.65 &  & 1.000 & 0.977 & 0.953 & 0.906 & 0.882 & 0.858 & 0.809 & 0.784 & 0.759 & 0.733 & 0.707 & 0.653 & 0.625 & 0.596 \\
0.70 &  & 1.000 & 0.973 & 0.945 & 0.888 & 0.859 & 0.830 & 0.770 & 0.739 & 0.707 & 0.675 & 0.641 & 0.570 & 0.533 &  \\
0.75 &  & 1.000 & 0.965 & 0.930 & 0.858 & 0.821 & 0.783 & 0.704 & 0.663 & 0.620 & 0.575 & 0.529 &  &  &  \\
0.80 &  & 1.000 & 0.958 & 0.915 & 0.826 & 0.780 & 0.733 & 0.634 & 0.581 & 0.527 &  &  &  &  &  \\
0.85 &  & 1.000 & 0.946 & 0.891 & 0.777 & 0.717 & 0.655 & 0.524 &  &  &  &  &  &  &  \\
0.90 &  & 1.000 & 0.910 & 0.818 & 0.624 & 0.522 &  &  &  &  &  &  &  &  &  \\ \hline
\end{tabular}
\caption{n=50}
\end{table*}

\begin{table*}[h!]
\begin{tabular}{clllllllllllllll}
\cline{1-1} \cline{3-16}
\begin{tabular}[c]{@{}c@{}}Class Balance \\ $k$\end{tabular} & \multicolumn{1}{c}{} & \multicolumn{1}{c}{} & \multicolumn{1}{c}{} & \multicolumn{1}{c}{} & \multicolumn{1}{c}{} & \multicolumn{1}{c}{} & \multicolumn{1}{c}{} & \multicolumn{2}{c}{error ($\epsilon$)} & \multicolumn{1}{c}{} & \multicolumn{1}{c}{} & \multicolumn{1}{c}{} & \multicolumn{1}{c}{} & \multicolumn{1}{c}{} & \multicolumn{1}{c}{} \\ \cline{1-1} \cline{3-16} 
 &  & 1.000 & 0.975 & 0.950 & 0.925 & 0.900 & 0.875 & 0.850 & 0.825 & 0.800 & 0.775 & 0.750 & 0.725 & 0.700 & 0.675 \\
0.50 &  & 1.000 & 0.974 & 0.949 & 0.923 & 0.898 & 0.872 & 0.846 & 0.821 & 0.795 & 0.769 & 0.742 & 0.716 & 0.689 & 0.662 \\
0.55 &  & 1.000 & 0.973 & 0.946 & 0.918 & 0.891 & 0.863 & 0.835 & 0.806 & 0.778 & 0.749 & 0.719 & 0.688 & 0.656 & 0.623 \\
0.60 &  & 1.000 & 0.970 & 0.940 & 0.909 & 0.878 & 0.846 & 0.814 & 0.781 & 0.747 & 0.712 & 0.676 & 0.637 & 0.596 & 0.551 \\
0.65 &  & 1.000 & 0.965 & 0.930 & 0.894 & 0.857 & 0.819 & 0.781 & 0.740 & 0.698 & 0.654 & 0.607 & 0.556 & 0.500 &  \\
0.70 &  & 1.000 & 0.958 & 0.915 & 0.871 & 0.825 & 0.778 & 0.729 & 0.677 & 0.622 & 0.564 & 0.500 &  &  &  \\
0.75 &  & 1.000 & 0.946 & 0.891 & 0.833 & 0.773 & 0.711 & 0.645 & 0.575 & 0.500 &  &  &  &  &  \\
0.80 &  & 1.000 & 0.926 & 0.849 & 0.768 & 0.684 & 0.595 & 0.500 &  &  &  &  &  &  &  \\
0.85 &  & 1.000 & 0.884 & 0.762 & 0.635 & 0.500 &  &  &  &  &  &  &  &  &  \\
0.90 &  & 1.000 & 0.889 & 0.763 & 0.631 & 0.502 &  &  &  &  &  &  &  &  &  \\ \hline
\end{tabular}
\caption{n=5000}
\end{table*}

\newpage
\subsection{Additional Examples of Risk Assessment Tools}

\begin{table*}[ht]
\centering
\small
\begin{tabular}[t]{| p{7em} | p{21em} | p{28em} |} 
 \hline
 \textbf{Domain} & \textbf{Tool Name} & \textbf{Examples of Reliance on AUC} \\ [0.5ex] 
 \hline 
 Criminal justice & Colorado Pretrial Assessment Tool (CPAT) and Colorado Pretrial Assessment Tool-Revised (CPAT-R)  & \textbullet \quad Model development (see \cite{Terranova_undated-pq}) \newline \textbullet \quad Model validation (see \cite{Terranova_undated-pq}) \newline \textbullet \quad ``Accuracy equity'' with regard to race, gender, and housing status (see \cite{Terranova_undated-pq}) \\
 \hline
 Criminal justice & Prisoner Assessment Tool Targeting Estimated Risk and Needs (PATTERN) & \textbullet \quad Model development (see \cite{PATTERNus2019first}) \newline \textbullet \quad Model revisions and validation (see \cite{pattern2021report}) \newline \textbullet \quad ``Racial neutrality'' (see \cite{pattern2020update, pattern2020revalidation, pattern2021report} ) \\
  \hline
Criminal justice & Correctional Offender Management Profiling for Alternative Sanctions (COMPAS) & \textbullet \quad Model validation and revisions (see \cite{Brennan2009-ph}) \newline \textbullet \quad ``Accuracy equity'' with regard to race (see \cite{dieterich2016compas}) \newline \textbullet \quad Further examples analyzed in \cite{fazel2022predictive} \\
 \hline
Criminal justice & Virginia Pretrial Risk Assessment Instrument (VPRAI) & \textbullet \quad Model validation and revisions (see \cite{lovins2015riverside}, \cite{danner2015risk} and \cite{danner2016race}) \\
 \hline
Criminal justice & Federal Pretrial Risk Assessment (PTRA) & \textbullet \quad Model development (see \cite{lowenkamp2009development}) \newline \textbullet \quad Model validation (see \cite{cohen2018revalidating}) \\
 \hline
  Criminal justice & El Paso Pretrial Risk Assessment
Instrument - Revised (EPPRA-R) & \textbullet \quad Model validation (see \cite{queen2022predictive})  \\
 \hline
 Child welfare & Allegheny Family Screening Tool (AFST) & \textbullet \quad Model development and revisions (see \cite{vaithianathan2017developing} and \cite{vaithianathan2019allegheny}) \newline \textbullet \quad Model fairness (see \cite{vaithianathan2017developing} and \cite{vaithianathan2019allegheny}) \\
 \hline
 Child welfare & Douglas County Decision Aide (DCDA) & \textbullet \quad Model development (see \cite{vaithianathan2019implementing}) \\
 \hline
 Child welfare & LA Risk Stratification Tool & \textbullet \quad Model development (see \cite{la_childwelfare_tool}) \newline \textbullet \quad Model fairness with regard to race (see \cite{la_childwelfare_tool})\\ 
 \hline
 Education & Wisconsin Dropout Early Warning System (DEWS) & \textbullet \quad Model development (see \cite{knowles2015wisconsinEWS}) \\
 \hline
\end{tabular}
\caption{Examples of Government Agencies Relying on AUC in Connection with Risk Assessment Tools in the criminal legal system, the child welfare system, and education.}
\label{tab:other-gov-examples}
\end{table*}%

\end{document}